\documentclass[pre,showpacs,twocolumn,preprintnumbers,amsmath,amssymb,superscriptaddress]{revtex4}
\usepackage{graphicx}
\usepackage{dcolumn}
\usepackage{bm}
\usepackage{natbib}
\usepackage[nativepdf,hyperfigures=false,colorlinks=true,citecolor=blue,linkcolor=black,anchorcolor=black,urlcolor=black]{hyperref}
\usepackage[absolute]{textpos}
\newcommand {\be}{\begin{equation}}
\newcommand {\ee}{\end{equation}}
\newcommand {\bea}{\begin{eqnarray}} 
\newcommand {\eea}{\end{eqnarray}}
\begin{document}
\title{Identifying phase synchronization clusters in spatially extended dynamical systems }
\author{Stephan \surname{Bialonski}}
\email{bialonski@gmx.net}
\affiliation{Department of Epileptology, Neurophysics Group, University of Bonn, Sigmund-Freud-Str. 25, D-53105 Bonn, Germany} 
\affiliation{Helmholtz-Institute for Radiation
and Nuclear Physics, University of Bonn, Nussallee 14-16, 53115 Bonn,
Germany}
\author{Klaus \surname{Lehnertz}}
\email{klaus.lehnertz@ukb.uni-bonn.de}
\affiliation{Department of Epileptology, Neurophysics Group, University of Bonn, Sigmund-Freud-Str. 25, D-53105 Bonn, Germany}
\affiliation{Helmholtz-Institute for Radiation and Nuclear Physics, University of Bonn, Nussallee 14-16, 53115 Bonn, Germany}
\affiliation{Interdisciplinary Center for Complex Systems, University of Bonn, R{\"o}merstr. 164, 53117 Bonn, Germany} 
\published{14 November 2006}

\begin{abstract}
We investigate two recently proposed multivariate time series analysis techniques that aim at detecting phase synchronization clusters in spatially extended, nonstationary systems with regard to field applications. The starting point of both techniques is a matrix whose entries are the mean phase coherence values measured between pairs of time series. The first method is a mean field approach which allows to define the strength of participation of a subsystem in a single synchronization cluster. The second method is based on an eigenvalue decomposition from which a participation index is derived that characterizes the degree of involvement of a subsystem within multiple synchronization clusters. Simulating multiple clusters within a lattice of coupled Lorenz oscillators we explore the limitations and pitfalls of both methods and demonstrate (a) that the mean field approach is relatively robust even in configurations where the single cluster assumption is not entirely fulfilled, and (b) that the eigenvalue decomposition approach correctly identifies the simulated clusters even for low coupling strengths. Using the eigenvalue decomposition approach we studied spatiotemporal synchronization clusters in long-lasting multichannel EEG recordings from epilepsy patients and obtained results that fully confirm findings from well established neurophysiological examination techniques. Multivariate time series analysis methods such as synchronization cluster analysis that account for nonlinearities in the data are expected to provide complementary information which allows to gain deeper insights into the collective dynamics of spatially extended complex systems. 
\end{abstract}
\maketitle

\begin{textblock*}{20cm}(3cm,27cm)
 Published as Phys. Rev. E \textbf{74}, 051909 (2006). Copyright 2006 by the American Physical Society.
\end{textblock*}

\section{Introduction}
Spatially extended complex dynamical systems may be thought of being composed of numerous constituents (dynamically formed subsystems) each having its own dynamics. Typically the relevant state variables of such systems can not be observed directly but only through some observation function that projects the high-dimensional state space onto an observation space of much lower dimension, resulting in a set of time series. Multivariate analyses of such time series might then help to gain deeper insights into the collective dynamics of spatially extended systems. Although a number of time series analysis methods have been developed over the past (see \cite{Brillinger81,Priestley88,BendatPiersol00,Pikovsky01,KS03} for an overview) most techniques either allow to characterize single time series (univariate approaches) or to investigate relationships between two time series (bivariate approaches). However, applying bivariate techniques to pairs of time series -- taken from a multichannel recording -- does not necessarily allow to identify the relevant information in the full data set. The latter is of particular interest for scientific fields investigating spatially extended dynamical systems, such as meteorology, economics, social science or neurosciences, where a complex but relatively sparse connectivity between subsystems prevails. Understanding brain function -- both during physiological and pathophysiological conditions (as e.g. in the case of epilepsy) -- requires a characterization and quantification of the collective behavior of neural networks generating signals at different areas. 

In principle, multivariate time series analysis techniques can be used to investigate mutual relationships between arbitrary numbers of time series. A large variety of methods \cite{Jain88} aim at revealing additional information by classifying time series into different groups. In addition to the classical principal component analysis (also known as Karhunen-Loeve transform) \cite{Mardia70} independent component analysis \cite{Hyv01} provides a decomposition of data into independent source signals, and if the assumption of independence holds, it can be regarded as a suitable method. If independence can not be assumed, mutual information based methods might be more appropriate \cite{Stoeg04,Kraskov05}. The partial coherence \cite{BendatPiersol00} measures the fraction of coherence between two time series that is not shared with a third time series. Whereas the partial coherence is based on the assumption of linearity and thus does not capture nonlinear interactions,
the recently proposed concept of partial phase synchronization \cite{schelter06} was designed to account for nonlinearities of the dynamics under investigation. In order to study {\em causal relations} among simultaneously acquired time series generated by linear stochastic systems the Granger causality \cite{Granger69} can be used by fitting autoregressive models. Besides a recently suggested nonlinear extension of Granger's ideas \cite{Chen04}, we mention the directed transfer function that is defined for an arbitrary number of channels \cite{Kaminski91} and is based on a multivariate autoregressive model approach \cite{Frana85}. In Refs. \cite{Samashima99,Baccala01} the partial directed coherence has been proposed for inference of linear Granger causality in the frequency domain based on vector autoregressive models of appropriate order. 
Both methods have been repeatedly applied to study interdependencies and causal relationships among neural signals (see e.g. \cite{Kaminski01,Hesse03,Blinowska04,Schelter06b} and references therein).

Over the last decade time series analysis techniques known from random matrix theory \cite{Brody81,Guhr98} have been repeatedly shown to allow an improved characterization of complex spatiotemporal correlation patterns. In these studies particularly the equal-time correlation matrix has been analyzed that was constructed from multivariate data sets obtained empirically in scientific fields ranging from economy \cite{Laloux99,Plerou99} and meteorology \cite{Santhanam01} to the neurosciences \cite{Kwapien00,Seba03,Mueller05a,Carmeli05,Mueller06}. More recently an approach for the detection of clusters in financial data based on properties of the eigenvalue spectrum of the equal-time correlation matrix was proposed in Ref. \cite{Kim05}. By filtering out the random part and collective market-wide effects the authors were able to detect groups of stocks by optimizing the matrix representation. Such an approach, however, requires a clear-cut definition of the random part, which is usually assumed to be associated with the small bulk eigenvalues. In Ref. \cite{Mueller05a} it was demonstrated though that the lower part of the eigenvalue spectrum may contain essential information.

Another way to study interactions in spatially extended systems is based on a statistical analysis of phase synchronization phenomena \cite{Pikovsky01,Bocca02}. In most studies, however, the analysis of empirical multivariate data has been accomplished by a repeated application of bivariate synchronization measures and it remains to be established whether this approach allows to fully characterize a common integrating structure that may be present in the data. Addressing this issue, Allefeld and Kurths proposed a genuinely multivariate phase synchronization analysis method \cite{Allefeld03,Allefeld04} and successfully applied their method to electroencephalographic (EEG) data recorded during a psychological experiment. The authors concluded their method to provide additional information on brain dynamics in a topographically, temporally, and frequency-specific way, as well as in other fields concerned with multivariate oscillatory processes. More recently, Allefeld and colleagues \cite{Allefeld05a} introduced a new approach that addresses a limitation of their original method, namely the assumption of the existence of a single synchronization cluster in the data. By using methods from random matrix theory the resulting approach appears to be capable of identifying multiple synchronization clusters in the data which makes it highly attractive to characterize pathophysiological, spatiotemporal synchronization phenomena in multichannel EEG recordings from epilepsy patients.

Recent findings indicate that bivariate analysis techniques allow to characterize physiological and pathophysiological phenomena in the human brain (see Ref. \cite{Stam2005,Mormann2006} for an overview), and it can be expected that multivariate phase synchronization analysis techniques provide complementary information. To address this issue we here study the synchronization cluster analysis methods proposed in Refs. \cite{Allefeld04,Allefeld05a} particularly with respect to field applications using model systems. In addition, we show that the method proposed in Ref. \cite{Allefeld05a} allows to detect multiple synchronization clusters in long-lasting multichannel EEG recordings from epilepsy patients.

This paper is organized as follows. Since both methods are based on the mean phase coherence we first recall its definition and interpretation as a bivariate measure for phase synchronization (Sec. \ref{sec:phase_synchronization}). In Sec. \ref{sec:synchronization_cluster_analysis} we briefly introduce the multivariate synchronization cluster methods, namely the mean field approach (Sec. \ref{sec:mean_field_approach}) and the eigenvalue decomposition approach (Sec. \ref{sec:eigenvalue_decomposition}). Next, we present our simulation studies aiming at an exploration of the limitations and a comparison of both methods (Sec. \ref{sec:simulations}). Finally, in Sec. \ref{sec:eeg-data} we present findings that were obtained from a spatiotemporal synchronization cluster analysis of multichannel EEG data recorded from epilepsy patients using the eigenvalue decomposition approach.

\section{Methods}
\subsection{Measuring phase synchronization}
\label{sec:phase_synchronization}
Phase synchronization was first described by Christiaan Huygens \cite{Hugenii73} in the 17th century and can be defined as the locking of the phases of two oscillating systems $j$ and $k$:
\be
\Delta\varphi_{jk}(t) = \varphi_j(t) -  \varphi_k(t) = const.
\ee
In a statistical way the degree of phase synchronization can be quantified by measuring the phase differences $\Delta\varphi_{jk}$ $n$ times and transforming them onto a unit cycle in the complex plane. The underlying circular distribution of the sample can be characterized by means of directional statistics \cite{Mardia00} with the \textit{mean phase coherence} $R_{jk}$ \cite{Hoke88,Mormann00}
\be
\label{eq:meanphasecoherence}
R_{jk} = \left| \frac{1}{n} \sum_{m=0}^{n-1} e^{i(\varphi_{jm}-\varphi_{km})}\right| 
\ee
where $\varphi_{jm}$ denotes the phase of system $j$ in measurement $m$. By definition, $R_{jk}$ is confined to the interval [0,1] where $R_{jk}=1$ indicates fully synchronized systems. In field applications the sample size $n$ is typically limited making $R_{jk}$ an estimate of the true {\em population value} $\rho_{jk}=\left|\left< e^{i(\varphi_j - \varphi_k)} \right> \right|$ of the underlying distribution of phase differences (the angle brackets denote the average over all members of the population).

When analyzing real valued time series $s(t)$ different methods can be used to extract phase information. Methods based on the Fourier, the Hilbert, or the wavelet transform were shown to be equivalent under relatively general assumptions \cite{Quiroga02,Bruns04}. The main idea is to map the data onto the complex plane using a function $z$ and to take the complex argument in order to obtain the phase $\varphi = \arg(z(t))$. We here followed an approach based on the Hilbert transform using the \textit{analytic signal} \cite{Gabor46,Panter65}
\be
z(t) = s(t) + i Hs(t)
\ee
where $Hs(t)$ denotes the Hilbert transform of the signal $s(t)$
\be
Hs(t) =\frac{1}{\pi}{\cal P}\int^{+\infty}_{-\infty}\frac{s(\tau)}{t-\tau}d\tau
\ee
and $\cal{P}$ denoting the Cauchy principal value of the integral. Application of the convolution theorem turns the last equation into 
\be
H s(t) = -i \mathcal{F}^{-1}\Big{[}\mathcal{F}[s(t)] \cdot sgn(\omega)\Big{]}
\ee
where $\mathcal{F}$ denotes the Fourier transformation and $\mathcal{F}^{-1}$ the inverse transformation, respectively. Thus, the imaginary part of the analytic signal is obtained by shifting each frequency component of the original signal by $\pi/2$ separately. It is important to keep in mind that phases may not have a physical meaning for arbitrary signals. If, however, the dynamics exhibit oscillations with a single main rhythm, then the phases are typically well defined \cite{Boashash92,Rosenblum-Mossbook01}.

In our applications (see Sec. \ref{sec:appl}) we computed $R_{jk}$ from the discrete time series $s_j(t)$ and $s_k(t)$ as follows. First, the data were normalized to zero mean, which corresponds to setting the DC Fourier coefficient to zero. Second, in order to avoid edge effects the first and last 10\% of a data window of size $n^\prime$ were tapered using a cosine half wave (Hanning window) before performing the Fourier transform. Third, since the computation of the Hilbert transform requires integration over infinite time, which cannot be performed for a window of finite length, 10\% of the calculated phase time series $\varphi_j(t)$ and $\varphi_k(t)$ were discarded on each side of the window, reducing the size of the considered phase time series to $n = 0.8 \cdot n^\prime$. For a data window of size $n$ and $m\in\{0,\ldots,n-1\}$ denoting the data point within the window the mean phase coherence $R_{jk}$ between the two time series with sampling interval $\Delta t$ was obtained by identifying $\varphi_{jm} = \varphi_j(m\Delta t)$ and applying Eq. (\ref{eq:meanphasecoherence}). 

\subsection{Synchronization cluster analysis}
\label{sec:synchronization_cluster_analysis}
For $N$ oscillating systems the pairwise computation of the mean phase coherence $R_{jk}$ leads to a matrix $\mathbf{R}$ which is symmetric due to definition (\ref{eq:meanphasecoherence}). Subsets of oscillating systems can be interpreted as {\em synchronization clusters} if these systems exhibit higher mean phase coherence values between each other than with systems not included in the same subset. Allefeld and colleagues \cite{Allefeld04,Allefeld05a} recently proposed two different multivariate approaches that aim at
identifying synchronization clusters in which the oscillating systems participate with different strength.
Both approaches are based on the bivariate mean phase coherence and shall briefly be recalled here.

\subsubsection{Mean field approach}
\label{sec:mean_field_approach}
In Ref. \cite{Allefeld04} a mean field approach has been presented which assumes the existence of a single synchronization cluster $C$ in the data. Generating a common rhythm all oscillating systems constitute the cluster but contribute to its emergence to a different extent. A mean field with phase $\Phi$ can then characterize the dynamics of the collective behavior. Using Eq. (\ref{eq:meanphasecoherence}) the degree of participation (\textit{participation strength}) of each oscillating system $j$ to the cluster $C$ can be quantified by
\be
\label{eq:partstrength}
R_{jC} = \left| \frac{1}{n} \sum_{m=0}^{n-1} e^{i(\varphi_{jm} - \Phi_m)}\right|\mbox{.}
\ee

A straightforward derivation of the phase $\Phi$ has been demonstrated in Ref. \cite{Allefeld04} for a simple model system. This method, however, requires exact knowledge about the underlying equations of motion and thus can not be applied to unknown systems in general. Nevertheless, the participation strength (\ref{eq:partstrength}) can be estimated directly as follows.

If the assumption of the existence of a single cluster holds, a mean field can be introduced such that the dynamics of the phase differences are decoupled. If, in addition, the noise affecting the oscillating systems is statistically independent for each system the phase differences $\Delta\varphi_j = \varphi_j - \Phi$ become independent random variables. Hence the population values $\rho_{jk}$ of the mean phase coherences $R_{jk}$ turn into
\bea
\nonumber
\rho_{jk} &=& \big{\vert} \big{<} e^{i(\varphi_j - \varphi_k)}\big{>} \big{\vert} 
= \big{\vert} \big{<} e^{i(\Delta\varphi_j - \Delta\varphi_k)}\big{>}\big{\vert} \\\nonumber
&=& \big{\vert} \big{<} e^{i\Delta\varphi_j} \big{>} \big{\vert} \cdot 
\big{\vert} \big{<} e^{-i\Delta\varphi_k} \big{>} \big{\vert} \\
&=& \rho_{jC} \cdot \rho_{kC}\;\mbox{ for }\;j\neq k\quad\mbox{ (}\rho_{jj} = 1\mbox{).}
\label{eq:factorization}
\eea
Taking into account that the mean phase coherence is asymptotically normally distributed $R_{jk} \sim {\cal N}(\rho_{jk},\sigma_{jk})$ \cite{Allefeld04b} and is an empirical estimate of $\rho_{jk}=\rho_{jC}\,\rho_{kC}$ (see Sect. \ref{sec:phase_synchronization}), a maximum likelihood estimation of the $\rho_{jC}$ leads to the minimization of the sum of squared weighted errors that defines the cost function 
\be
\label{eq:e_sum}
\Gamma = \sum_{j>k} \gamma_{jk}^2 \quad \mbox{with}\quad \gamma_{jk} = \frac{R_{jk}-\rho_{jC}\,\rho_{kC}}{\sigma_{jk}}\mbox{.}
\ee
Assuming that the circular distribution of the phase differences can be sufficiently approximated by a wrapped normal distribution, the standard deviation $\sigma_{jk}$ of the sampling distribution of $R_{jk}$ can be expressed in terms of $\rho_{jk}$ (cf. \cite{Allefeld04b,Mardia00})
\be
\label{eq:standard_dev_sigma}
\sigma_{jk} = \frac{1}{\sqrt{2n}} (1-\rho_{jC}^2\,\rho_{kC}^2)\mbox{.}
\ee
In the following we refer to $\rho_{jC}$ as defined above as $R_{jC}$, thereby emphasizing an interpretation as a {\em to-cluster synchronization strength} analogous to Eq. (\ref{eq:partstrength}). In our applications participation strengths were computed by minimizing $\Gamma$ using an iterative algorithm proposed in Ref. \cite{Allefeld06a}.

\subsubsection{Eigenvalue decomposition approach}
\label{sec:eigenvalue_decomposition}
In Ref. \cite{Allefeld05a} another approach has recently been proposed that is based on the eigenvalue decomposition of the matrix $\mathbf{R}$ and appears to allow identification of multiple clusters. The procedure makes use of findings by M\"uller and colleagues \cite{Mueller05a} who demonstrated that information about the correlation structure of multivariate data sets is imprinted into the dynamics of the eigenvalues and into the structure of the corresponding eigenvectors by nonrandom level repulsion.

The eigenvalues $\lambda_c$ and eigenvectors $\vec{\nu}_c$ of the symmetric and real-valued matrix $\mathbf{R}$ (see Eq. \ref{eq:meanphasecoherence}) can be obtained by solving the eigenvalue equation 
\be
\mathbf{R} \cdot \vec{\nu}_c = \lambda_c \cdot \vec{\nu}_c, \quad c \in \{1,\ldots,N\}
\ee
which, in general, has $N$ different solutions. In the following it is assumed, that 
the eigenvectors are normalized ($|\vec{\nu}_c | = 1$). Being transformed by an orthogonal transformation into the basis of its eigenvectors $\mathbf{R}$ becomes the diagonal matrix $\mathbf{D}$ of its eigenvalues. The invariance of the trace under this transformation leads to the equation $N = \mbox{tr}(\mathbf{R}) = \mbox{tr}(\mathbf{D}) = \sum_{c}\lambda_c$. In the case of systems $j,k$ showing no phase synchronization ($R_{jk} = 0$ for $j\neq k$) the equation is trivially fulfilled by $\lambda_c = 1$ for all $c$, whereas the occurrence of entries $R_{jk} > 0$ for $j \neq k$ induces a level repulsion, a combined increase and decrease of eigenvalues in such a way that $N = \sum_{c}\lambda_c$ still holds.

A multivariate analysis is realized as follows (cf. Ref. \cite{Allefeld05a}): Each eigenvalue $\lambda_c > 1$ is associated with a synchronization cluster and quantifies its strength within the data set. The internal structure of cluster $c$ is described by the corresponding eigenvector $\vec{\nu}_c$. Being normalized ($\sum_j\nu_{jc}^2 = 1$) its components quantify the relative involvement of each system $j$ to cluster $c$ by $\nu_{jc}^2$. Combining the eigenvalue $\lambda_c$ and the index $\nu_{jc}^2$ the ``absolute" involvement of a system $j$ in a cluster $c$ can be described by the \textit{participation index} 
\be
p_{j,c} = \lambda_c \cdot \nu_{jc}^2\mbox{.}
\label{eq:particip}
\ee
Consequently, system $j$ is considered as belonging to cluster $c$ for which its participation index becomes maximal.

In case of non-vanishing entries of $\mathbf{R}$ between systems belonging to different clusters (inter-cluster synchronization) it can be observed that these clusters are not characterized by separate but by a superposition of eigenvectors. In order to adjust the interpretation of the participation indices as the degree of involvement of a system within one cluster (see. Eq. \ref{eq:particip}) it was proposed in Ref. \cite{Allefeld05a} to compute the $p_{j,c}$ in a first step and to assign the systems to the clusters as mentioned above. In a second step the matrix entries representing inter-cluster synchronization are set to zero and the participation indices are computed on the trimmed matrix again. 
In our applications we followed exactly this scheme. 

Summarizing this section we conclude that both methods, the mean field and the eigenvalue decomposition approach, seem to provide similar results in single cluster configurations for which an almost functional dependency of $R_{jC}^2 = p_{j,1}$ ($\lambda_1$ denoting the largest eigenvalue) can be observed \cite{Allefeld05a}. Nevertheless, the question whether the approaches provide meaningful results in case of multi-cluster configurations -- particularly with regard to field applications -- remains unaddressed and shall be investigated in the following section. 

\section{Applications}
\label{sec:appl}
\subsection{Simulations}
\label{sec:simulations}
\begin{figure}[bh]
\includegraphics[width=65mm]{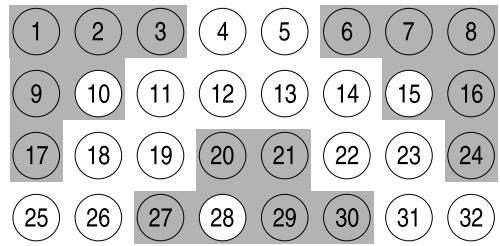}
\caption{Three-cluster configuration consisting of a lattice of 32 coupled Lorenz systems. Systems within a gray-shaded area belong to one cluster and are driven by one of the driving systems $D_1$ = 10, $D_2$ = 15 and $D_3$ = 28 which are highlighted in white.}
\label{fig:lorenz-config}
\end{figure}

We here studied a three-cluster configuration which consisted of a lattice of 32 coupled identical Lorenz systems (cf. Fig. \ref{fig:lorenz-config}).
Each system $j$ is defined by the differential equations
\bea
\dot{x}_j &=& -\frac{8}{3} \cdot x_j + y_j z_j +\epsilon_j \cdot (x_{D_{1,2,3}}-x_j)\\\nonumber
\dot{y}_j &=& 28 \cdot z_j - y_j -x_jz_j\\\nonumber
\dot{z}_j &=& 10 \cdot (y_j-z_j)
\eea
where $\epsilon_j$ denotes the coupling strength controlling the influence of the diffusive unidirectional coupling on system $j$ by one of three chosen driving systems $D_{1,2,3}$.

\begin{figure*}[thb]
\includegraphics{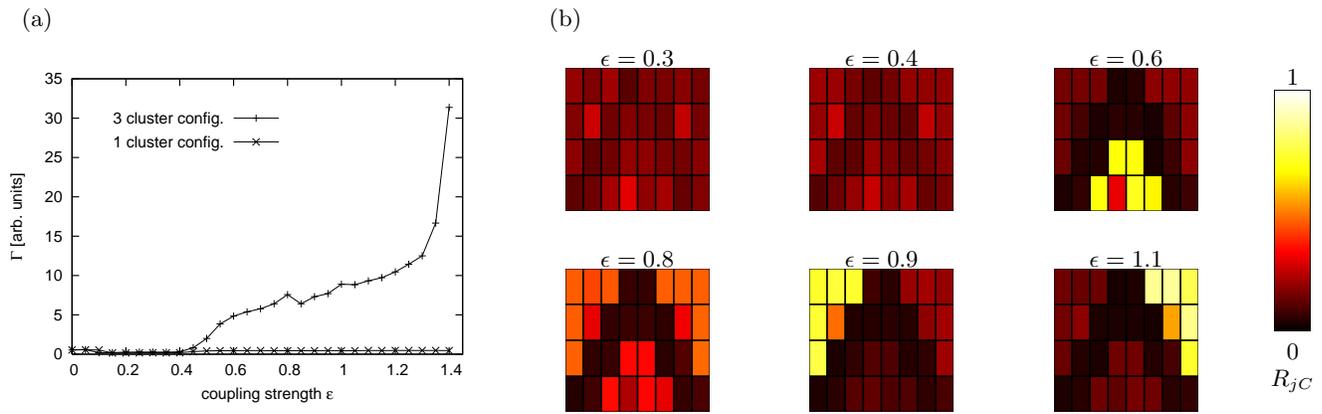}
\caption{(Color online) (a) Dependence of the cost function $\Gamma$ on the coupling strength $\epsilon$ for a single-cluster and a three-cluster configuration. (b) Spatial distribution of color-coded participation strength $R_{jC}$ for selected values of $\epsilon$ (see Fig. \ref{fig:lorenz-config} for arrangement of clusters).}
\label{fig:lorenz-E-part-strength}
\end{figure*}

\begin{figure}[bht]
\includegraphics{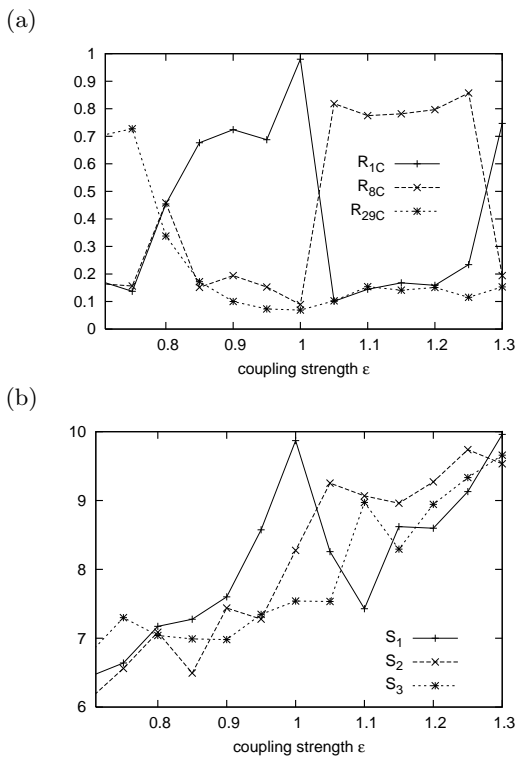}
\caption{(a) Dependence of participation strength $R_{jC}$ on the coupling strength $\epsilon$ for exemplary systems $j \in \{1,8,29\}$. (b) Dependence of cluster strength $S_d, d=1,\dots, 3$ on the coupling strength $\epsilon$.}
\label{fig:lorenz-cluster-strength}
\end{figure}

In order to control this configuration by a single parameter we set $\epsilon_j = \epsilon$ for systems within each cluster while $\epsilon_j = 0$ for the uncoupled systems. Taking randomly chosen points in the state space near the Lorenz attractor as initial conditions the differential equations were iterated using a fourth order Runge-Kutta algorithm \cite{Press92} with a step size of 0.01. In order to eliminate transients, the first $10^4$ iterations were discarded. For increasing coupling strengths $\epsilon = 0.0$ to $1.4$ (step size $0.05$) we generated scalar time series of the $x$-components ($n^\prime = 5\cdot 10^5$ data points) and computed the participation strength $R_{jC}$ and the participation indices $p_{j,c}$. 

For the mean field approach the configuration represents a violation of the single-cluster assumption.
When increasing the coupling strength $\epsilon$ the mean phase coherence values between the coupled systems increased and consequently the three synchronization clusters emerged. This is reflected by the dependency of the cost function $\Gamma$ on $\epsilon$ as shown in Fig. \ref{fig:lorenz-E-part-strength}a. Since the single-cluster assumption leading to Eq. (\ref{eq:factorization}) was violated, $\Gamma$ increased rapidly for $\epsilon > 0.4$. For comparison, we repeated the analysis for a single-cluster configuration that was generated by setting the coupling terms between the systems of clusters driven by $D_1$ and $D_2$ to zero. Here $\Gamma \le 0.6$ for all $\epsilon$ values (cf. Fig. \ref{fig:lorenz-E-part-strength}a). Using $\Gamma$ as an indicator that reflects the violation of the single-cluster assumption (cf. Ref. \cite{Allefeld05a}), which increases with the internal degree of synchronization between systems within a cluster, we generated plots 
(cf. Fig. \ref{fig:lorenz-E-part-strength}b) showing the spatial distribution of the participation strength $R_{jC}$ for the three-cluster configuration chosen here. For $\epsilon \le 0.4$ ($\Gamma \le 0.6$) no clear-cut cluster structure could be identified in the spatial distribution of participation strengths. For $\epsilon \geq 0.5$ ($\Gamma \gg 0.6$) either a single or all three synchronization clusters emerged, however, with a varying degree of visibility. Interestingly, for various coupling strengths (Fig. \ref{fig:lorenz-E-part-strength}b, e.g. $\epsilon \in \{0.6,0.9,1.1\}$) the participation strengths of systems involved in one cluster exhibited higher values than the remaining participation strengths. This behavior is shown in more detail in Fig. \ref{fig:lorenz-cluster-strength}a for exemplary systems involved in one cluster. In order to further elucidate this phenomenon we calculated the sum $S_d = \sum_{j>k; j,k \in C_d} R_{jk}$, where $C_d$ denotes the index set of systems belonging to cluster $d$. $S_d$ quantifies the strength of cluster $d$ and increased, on average, with an increasing coupling strength $\epsilon$ (cf. Fig. \ref{fig:lorenz-cluster-strength}b) while fluctuations can be attributed to the mean phase coherence values being asymptotically normally distributed. A synchronization cluster $d$ became visible in the spatial distribution of
participation strengths $R_{jC}$ when its strength $S_d$ dominated the configuration (cf. Fig. \ref{fig:lorenz-cluster-strength}). This effect was caused by the cost function $\Gamma$ exhibiting several local minima whose values changed according to the cluster strengths $S_d$. For $\epsilon = 0.8$ all $S_d$ attained similar values. Here the minimization algorithm terminated because of having found a single minimum which would lead to the (mis)interpretation of all systems belonging to one single cluster. 

\begin{figure}[tb]
\includegraphics{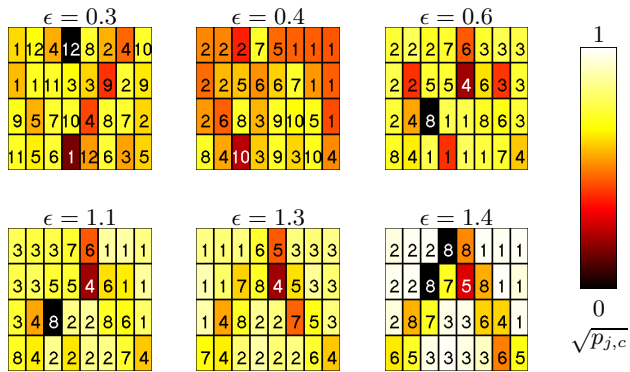}
\caption{(Color online) Spatial distribution of color-coded participation indices $p_{j,c}$ for selected values of the coupling strength $\epsilon$. Numbers denote the clusters $c$ to which each system was assigned by the eigenvalue decomposition method. Eigenvalues were sorted in a descending order.}
\label{fig:lorenz_systems_multiclusteranalysis}
\end{figure}

In contrast to the mean field method, the eigenvalue decomposition approach provides not just a single set of scalars but multiple sets of scalars representing different clusters. The eigenvalues were labeled here according to $\lambda_1 \geq \ldots \geq \lambda_N$ thereby enabling an easy identification of the dominating cluster structures by the corresponding index $c$. As shown in Fig. \ref{fig:lorenz_systems_multiclusteranalysis} for low coupling strengths ($\epsilon \leq 0.3$) the method was not able to detect the three different clusters but indicated a multitude of small clusters. This can be attributed to random fluctuations of the mean phase coherences. Increasing slightly the coupling strength to $\epsilon = 0.4$ two of the three simulated clusters were already detected in the two largest eigenvalues $\lambda_1$ and $\lambda_2$ whereas the corresponding participation strengths did not reveal any cluster structure (cf. Fig. \ref{fig:lorenz-E-part-strength}b). For $\epsilon \geq 0.5$ all three clusters were clearly visible and identified by the three largest eigenvalues. Their participation indices increased with increasing coupling strength $\epsilon$ (see e.g. $\epsilon = 1.1$ to $1.4$) thereby reflecting the internal degree of synchronization within the clusters. The cluster indices $c$ changed for different coupling strengths $\epsilon$ due to the fact that the cluster strengths fluctuated (cf. $S_d$ in Fig. \ref{fig:lorenz-cluster-strength}b) which is reflected by the eigenvalues (see e.g. $\epsilon \in \{1.1,1.3\}$ in Fig. \ref{fig:lorenz-cluster-strength}b and Fig. \ref{fig:lorenz_systems_multiclusteranalysis}).

In contrast to the mean field approach the eigenvalue decomposition approach was capable of distinguishing between different clusters (see \cite{Allefeld05a}). Limitations of the method, however, that are related to parameters influencing the detectability of different clusters have not been studied so far. To address this issue we designed a configuration consisting of $N = 32$ systems that formed two clusters $C_1= \{1,\ldots,r\}$ and $C_2 = \{r+1,\ldots,N\}$ where the parameter $r$ controls the relative size of an individual cluster. We did not study dynamical systems here but instead assigned values to the entries of matrix $\mathbf{R}$ that were drawn from normal distributions given that the mean phase coherences are asymptotically normally distributed (see section \ref{sec:mean_field_approach}). Mean phase coherences between systems of different clusters were drawn from the inter-cluster distribution ${\cal N}(\rho^{(int)},\sigma^{(int)})$, while those representing the entries between systems of the same cluster $C_1$ and $C_2$ were drawn from ${\cal N}(\rho^{(1)},\sigma^{(1)})$ and ${\cal N}(\rho^{(2)},\sigma^{(2)})$, respectively. For a given population value $\rho$ the standard deviation $\sigma$ was determined by equation (\ref{eq:standard_dev_sigma}) using a sample size of $n=200$.

Recalling that the detection of different clusters is based on the participation indices computed in the first processing step of the method (see Sec. \ref{sec:eigenvalue_decomposition}), which assigns a system to a cluster for which its participation index becomes maximal, we here quantify an erroneous assignment by considering the differences of the participation indices obtained from the first processing step $\chi_{j12} = p_{j,1} - p_{j,2}$ of system $j$ and clusters $C_1$ and $C_2$. For the configuration considered here we define the weighted amount of erroneous assignments as 
\be
\Delta = \left\{
\begin{array}{r@{\quad:\quad}l}
\Delta_1 + \Delta_2 & N_{\lambda} = 2 \\
0 & \mbox{else}
\end{array} \right.
\label{eq:delta}
\ee
where $N_{\lambda}$ denotes the number of eigenvalues $\lambda > 1$ and $\Delta_1$ and $\Delta_2$ are determined by
\bea
\Delta_1 &=& \left( \sum_{j = 1}^r\Theta(\chi_{j21}) \right)^{-1} \sum_{j = 1}^r \chi_{j21} \cdot\Theta(\chi_{j21}) \\\nonumber
\Delta_2 &=& \left(\sum_{j = r+1}^{N}\Theta(\chi_{j12})\right)^{-1} \sum_{j = r+1}^{N} \chi_{j12} \cdot \Theta(\chi_{j12})\mbox{,}
\eea
where $\Theta$ is the Heaviside step function. If system $j$ is erroneously assigned to the cluster $C_1$ but belongs to a cluster $C_2$ by construction, $\Theta(\chi_{j12})$ will be larger than zero. Thus $\Delta_2$ ($\Delta_1$) sums up incorrect assignments of systems $j$ to cluster $C_1$ ($C_2$) weighted by the differences of the corresponding participation indices $\chi$. If all systems are assigned to the correct cluster, then $\Delta=0$ by definition. Since the eigenvalues quantify the cluster strengths, a variation of the relative size of an individual cluster (by varying $r$) makes the largest eigenvalue traverse a minimum. This enables us to label the eigenvalues $\lambda_c > 1, c \in \{1,2\}$ according to their clusters $C_1$ or $C_2$.

\begin{figure}[tb]
\includegraphics[width=85mm,bb=50 70 410 285]{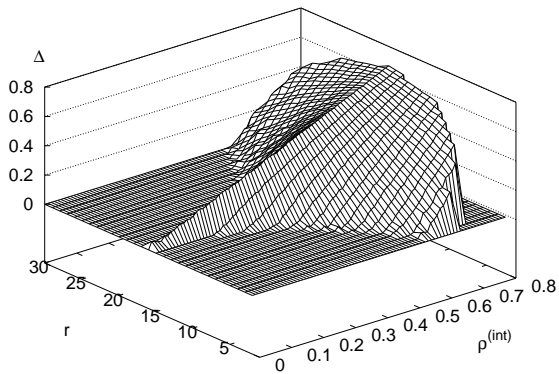}
\caption{Dependence of the error $\Delta$ on the relative size $r$ of an individual cluster and on the inter-cluster synchronization level $\rho^{(int)}$.}
\label{fig:delta_mui}
\end{figure}

We studied a transition from a configuration that consisted of two clusters $C_1$ and $C_2$ ($\rho^{(1)} = \rho^{(2)} = 0.8$) to a single-cluster configuration $C_0 = \{1,\ldots,N\}$ by successively increasing $\rho^{(int)}$ from 0.0 to 0.8 in steps of 0.01. Fig. \ref{fig:delta_mui} shows the weighted amount of erroneous assignments $\Delta$ depending on the relative size of an individual cluster $r$ and on the inter-cluster synchronization level $\rho^{(int)}$. The eigenvalue decomposition method successfully identified ($\Delta = 0$) the two clusters for low values of the inter-cluster synchronization level ($\rho^{(int)} < 0.17$) except for clusters of equal size ($r = N/2$). The weighted amount of erroneous assignments rapidly increased, however, for increasing $\rho^{(int)}$, thereby failing to detect the cluster structures for an increasing set of $r$ values. The number of eigenvalues $N_{\lambda}$ being larger than 1 appeared to be a sensitive measure for the number of clusters present in the data. We observed $N_{\lambda} \rightarrow 1$ (i.e., $\Delta=0$ by definition, see Eq. \ref{eq:delta}) only for relatively large values of the inter-cluster synchronization level ($\rho^{(int)} > 0.7$) where the transition to a single-cluster configuration was completed.

\begin{figure}[tb]
\includegraphics{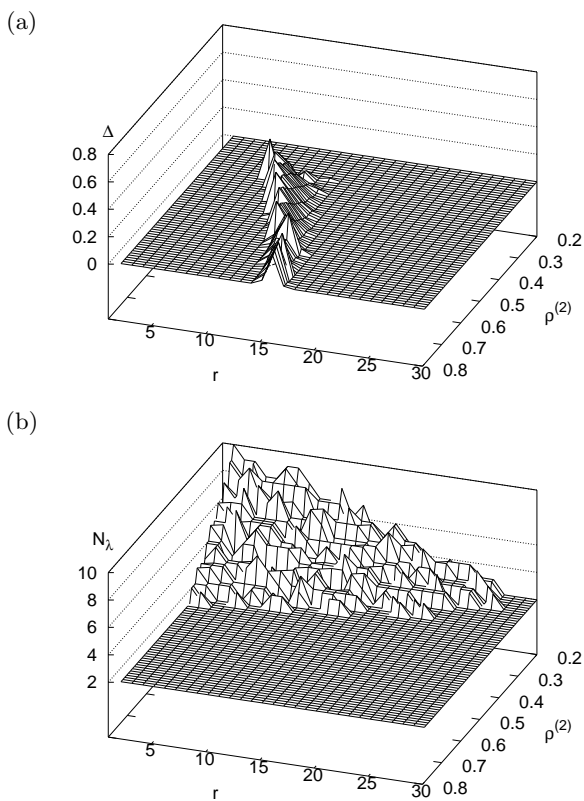}\\
\caption{Dependence of the error $\Delta$ (a) and $N_{\lambda}$ (b) on the relative cluster size $r$ and on the mean cluster synchronization level $\rho^{(2)}$ of cluster $C_2$.}
\label{fig:delta_mu2}
\end{figure}

Next we studied a transition from a configuration of two clusters $C_1$ and $C_2$ ($\rho^{(1)} = \rho^{(2)} = 0.8$, $\rho^{(int)} = 0.2$) to a configuration with only one cluster (here $C_1$) by decreasing $\rho^{(2)}$ in steps of 0.01 down to 0.2 (see Fig. \ref{fig:delta_mu2}a). The weighted amount of erroneous assignments $\Delta$ vanished for $\rho^{(2)}$ approaching the mean synchronization level of the inter-cluster distribution ($\rho^{(int)} = 0.2$). In this range the number of eigenvalues $N_{\lambda}$ increased (Fig. \ref{fig:delta_mu2}b). These additional structures were formed by those systems, which were previously involved in cluster $C_2$ for higher values of $\rho^{(2)}$. This fragmentation of a cluster $c$ can also be observed when increasing the corresponding standard deviation $\sigma^{(c)}$ independently from the population value $\rho^{(c)}$, thereby taking into account various uncertainties (e.g. definition of observables, measurement precision, finite sample size, phase extraction). Since the additional structures were caused by random fluctuations of the mean phase coherences within the cluster, we refer to them as \textit{pseudo-clusters} in the following. In our simulations we observed pseudo-clusters of different sizes, whose corresponding eigenvalues were only slightly larger than unity compared to the eigenvalues of the true clusters.

\begin{figure}[t]
\includegraphics[width=80mm]{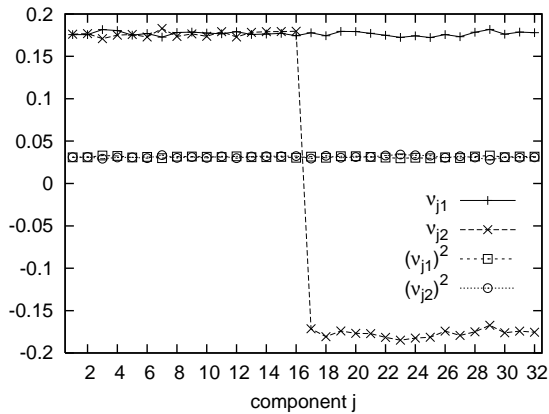}
\caption{Components $j$ of eigenvectors belonging to the two largest eigenvalues $\lambda_1$ and $\lambda_2$ for the case of $r = 16$, $\rho^{(1)} = \rho^{(2)} = 0.8$, $\rho^{(int)} = 0.2$.}
\label{fig:eigenvector_components}
\end{figure}

When varying $\rho^{(2)}$ the set of $r$ values for which $\Delta > 0$ remained almost stable but was shifted to lower $r$ values (Fig. \ref{fig:delta_mu2}a). 
Therefore, together with the relative size of an individual cluster the level of synchronization between systems within a cluster must be regarded as crucial for the eigenvalue decomposition approach. The results shown in Figs. \ref{fig:delta_mui} and \ref{fig:delta_mu2}a can then be sufficiently explained with the help of the cluster strength $S_d$ which considers both, the cluster size and the level of synchronization between systems within a cluster. The method failed to correctly assign all systems to their clusters in case of similar cluster strengths $|S_1 - S_2| < \kappa$, where $\kappa$ increases with an increasing level of inter-cluster synchronization. The latter leads to superpositions in the eigenvector components and, as a result, makes it impossible for the method to correctly detect the clusters by the maximum participation index criterion in certain configurations. In order to further elucidate this phenomenon we show in Fig. \ref{fig:eigenvector_components} the components $\nu_{jc}$ of the eigenvectors corresponding to the two largest eigenvalues for the case $r=16$ and $\rho^{(2)} = 0.8$. The squared components $\nu_{jc}^2$ of both vectors attained similar values, causing the method to classify all systems as belonging to a single cluster. Nevertheless, the two clusters $C_1$ and $C_2$ were clearly visible in the linear combinations $\vec{\nu}_1 \pm \vec{\nu}_2$.

Summarizing this section we conclude that the eigenvalue decomposition approach successfully identified the three synchronization clusters in our simulations using coupled Lorenz systems and, moreover, appears to be sensitive to detect clusters even for low coupling strengths. Although the mean field approach is by definition not able to distinguish between different clusters, the method appears to be relatively robust even in situations where the single-cluster assumption is not (entirely) fulfilled, leading to higher participation strengths for systems of the dominating cluster in our simulation. When comparing both methods the eigenvalue decomposition approach can be regarded as superior when analyzing data exhibiting different synchronization clusters, which can be expected e.g. for multichannel EEG data. 

\subsection{EEG data}
\label{sec:eeg-data}
In this section we present exemplary findings obtained from applying the eigenvalue decomposition method to multichannel EEG time series that were recorded from three epilepsy patients (denoted as P1, P2, and P3) suffering from pharmacoresistant focal epilepsies of neocortical origin. For these patients complete seizure control can be obtained by resecting the part of the brain responsible for seizure generation (epileptic focus). This requires an exact localization of the epileptic focus and its delineation from functionally relevant brain structures during the presurgical workup. When no concordant information can be achieved from noninvasive diagnostic techniques, the EEG is recorded from implanted electrodes over a longer period, typically 2--3 weeks. The analyses reported here were made after surgery had taken place, and after it had become clear from its success whether the localization of the epileptic focus had been correctly predicted. All patients had signed informed consent that their clinical data might be used and published for research purposes, and the study was approved by the local medical ethics committee.

Previous studies have shown that even during seizure-free intervals the seizure generating area of the brain exhibited higher interdependences \cite{Arnhold99} and an higher degree of synchronization \cite{Mormann00} than other brain areas. Together with results obtained from applying univariate time series analysis techniques (see e.g. \cite{Andr06} and references therein) these findings allow an improved understanding of intermittent dysfunctioning of the brain between seizures and provide potentially useful diagnostic information. We here addressed the question whether complementary information can also be obtained from a multivariate approach. Specifically, we investigated whether the eigenvalue decomposition approach (a) allows to localize the epileptic focus analyzing EEG recordings from the seizure-free interval only, and (b) is capable of detecting short time changes of synchronization patterns associated with physiological processes in the human brain. To this end we analyzed continuous EEG recordings that lasted 39 h for patient P1 and 26 h for patient P2 covering different physiological and pathophysiological states of the patients. In addition, we analyzed an EEG recording lasting 30 minutes during which patient P3 was simultaneously presented item pairs, either word pairs or unpronounceable letter string pairs of 5-11 letters length, and was instructed to press a button if he/she recognized word pairs with identical or highly similar meanings (synonyms) (task T1) or identical letter strings (task T2). No button press was demanded in case of non-synonym word pairs or letter strings in which one consonant differed. Tasks alternated every 3 min. The whole sequence was adopted from Ref. \cite{Fernandez01} and comprised three blocks of word and letter string pairs each. A baseline recording of 10 minutes was performed with eyes open before and after the experiment. Recent findings \cite{Price00,Weber06} indicate that tasks such as T2 are associated with a greater demand for decision-making processes due to the involvement of different phenomena like reading, phonological retrieval, or orthographic analysis during such conditions. We thus hypothesized to observe a higher degree of collectivity among neuronal assemblies particularly in brain areas associated with language processing. We did not expect to observe synchronization phenomena associated with the epileptogenic process in these brain structures from patient P3 since the epileptic focus was localized in a more distant brain structure. 

\begin{figure}
\includegraphics[width=45mm]{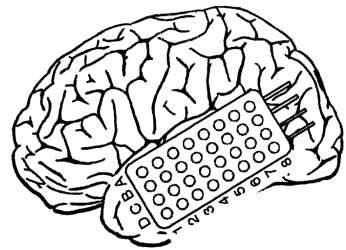}\\
\caption{Schematic view of the electrode grid with $N = 8\times 4$ contacts placed over the left temporal lateral neocortex.}
\label{fig:schema}
\end{figure}

\begin{figure}[tbh]
\includegraphics[width=85mm]{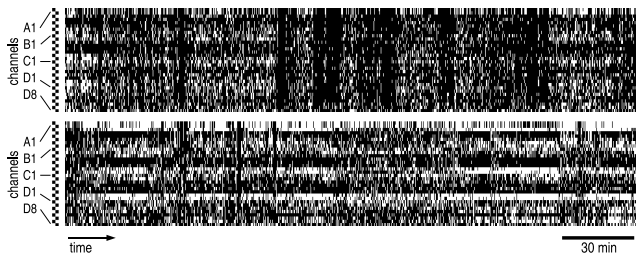}
\caption{Examples of the spatio-temporal evolution of the cluster defined by the largest eigenvalue. EEG data from patient P1 recorded during day-time (top) and during night-time (bottom). For each segment (duration: 16.38 s) channels belonging to the cluster are drawn in black. Note that channels A1 and A2 were used as reference during the recording.}
\label{fig:bit_strings_pat_P1}
\end{figure}

The EEG was measured from grid electrodes (rectangular flexible grids of $N = 8\times4$ contacts) placed onto the temporal lateral neocortex (see Fig. \ref{fig:schema}). EEG data were sampled at 200 Hz using a 16-bit analog-to-digital converter and filtered within a frequency band of 0.3--70 Hz. Using a moving-window technique EEG signals were divided into segments of $n^\prime = 4096$ sampling points each, and segments overlapped by 20\%. The length of the resulting segments corresponded to 16.38 s at the given sampling rate and can be regarded as a compromise between the required statistical accuracy for the calculation of the mean phase coherence and the approximate stationarity within a segments length (see Ref. \cite{Mormann03a} and references therein). After the calculation of the mean phase coherence (cf. Sect. \ref{sec:phase_synchronization}) for all channel combinations from each segment, the eigenvalue decomposition method was applied (cf. Sect. \ref{sec:eigenvalue_decomposition}). 

In order to estimate the number of clusters being detected on average in the data we sorted eigenvalues and corresponding eigenvectors of each segment in a descending order ($\lambda_1 \geq \ldots \geq \lambda_N$) and computed the mean values $\bar{\lambda}_c$ over all segments for each patient. The five to six largest averaged eigenvalues $\bar{\lambda}_c$ were larger than 1 causing the method to detect on average an according number of clusters per segment. Based on our findings presented in Sect. \ref{sec:simulations} we expected the lower part of the eigenvalue spectrum being larger than 1 to yield a considerable amount of pseudo-clusters. We therefore restricted further analyses to the three largest eigenvalues.

\begin{figure}[tb] 
\includegraphics[width=85mm]{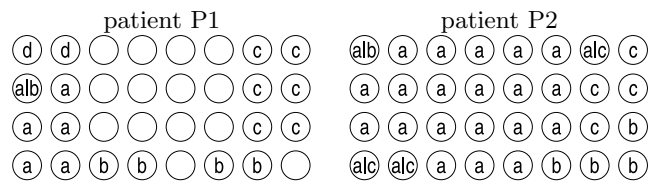}
\caption{Drawings of the grid with electrode contacts A1 (top left), A8 (top right) to D8 (bottom right). Channels participating in a cluster are marked by the same letter. Because of the averaging applied here a channel can be observed to participate in multiple clusters for different segments.}
\label{fig:bitstrings}
\end{figure}

Since a cluster is represented by a set of participating channels, each cluster can be written as a {\em bit string} of size $N$ where the bits represent the channels being involved (1) or not involved (0) in the cluster. This allowed us to discard the information about the relative involvement of each system within one cluster as reflected by the participation indices and to handle the clusters in a convenient way, namely by considering their simplified representation. Bit strings connected to eigenvalues fulfilling the threshold criterion ($\lambda > 1$) varied largely over time as shown exemplarily in Fig. \ref{fig:bit_strings_pat_P1}. Apart from frequently appearing bit strings in which all channels were involved, structures in time could be observed, where successive bit strings differed only in few bits. Given that the mean phase coherence fluctuates (due to measurement noise, a limited number of data points, or even due to short-term physiological or pathophysiological phenomena within a segment), the same cluster structure cannot be expected to show up in exactly the same bit string representation but may vary in some bits. In order to minimize this effect we sorted bit strings that were visible for more than 6 minutes in the data into groups which differed only in up to three bits. Moreover, we discarded groups of bit strings which represented clusters of size 0, 1, or 32 channels. In Fig. \ref{fig:bitstrings} we show groups that represented the largest number of bit strings in the EEG data. Interestingly, for both patients the spatial distribution of electrodes involved in the cluster labeled ``$a$'' corresponded to the spatial extent of the epileptic focus as determined by the presurgical workup. This cluster could be observed throughout the datasets and in total for at least 112 minutes for patient P1 and 125 minutes for patient P2. Indeed, the patients were operated on exactly this region and are now free of seizures. Thus, we conclude that the method seems to be quite sensitive for detecting synchronization clusters in EEG time series recorded from epilepsy patients even during the seizure-free interval.

\begin{figure*}[tbh]
\includegraphics[width=170mm]{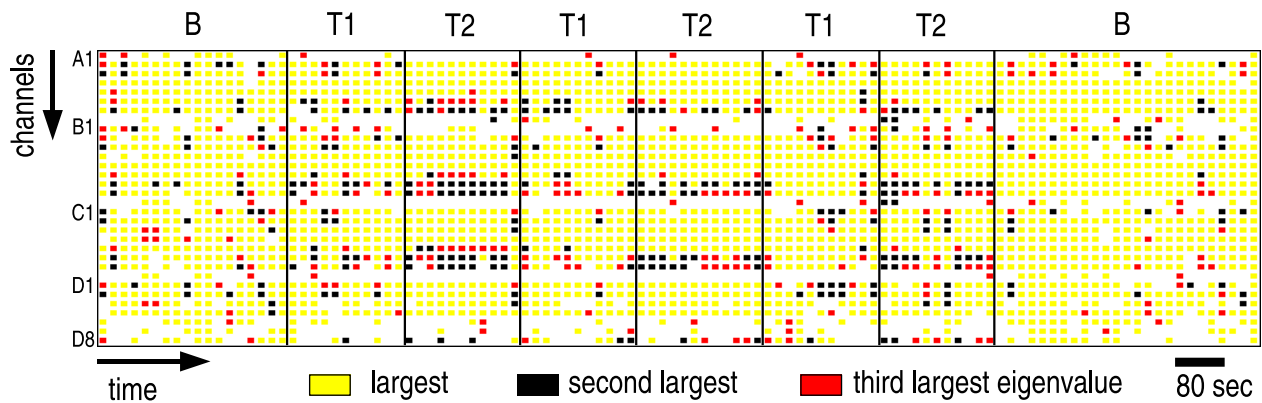}
\caption{(Color online) Spatio-temporal evolution of clusters defined by the largest, second-, and third-largest eigenvalues calculated from EEG data from patient P3 recorded during a language processing paradigm involving two different tasks (separated by black lines). The patient was simultaneously presented item pairs, either word pairs or unpronounceable letter string pairs, and was instructed to press a button upon recognizing synonyms (T1) or identical letter strings (T2). A baseline recording (B) was performed before and after the experiment. Each pixel in the figure represents the color-coded cluster membership of a given electrode contact (A1--A8, B1--B8, C1--C8, D1--D8 shown on the ordinate; cf. Fig. \ref{fig:schema})  and for a given EEG segment of duration 16.38 s (abscissa). Contacts A7, B7--8, and C7--8 covered a brain area associated with language processing (Wernicke's area) and were most noticeable during task T2.}
\label{fig:language_experiment}
\end{figure*}

Whereas the aforementioned results were obtained by analyzing the bit string groups which corresponds to a temporal average (over all segments), we now investigated whether the eigenvalue decomposition approach allows to detect short-term changes of cluster structures that can be related to physiological synchronization phenomena (the language processing paradigm mentioned above). The participation indices were computed from the EEG recording from patient P3 and translated into bit strings using the same criteria as mentioned above. Bit strings corresponding to the three largest eigenvalues are shown in Fig. \ref{fig:language_experiment}. Interestingly, a modulation of the bit strings depending on the tasks can be observed. The occurrence of bit strings representing a cluster of the electrode contacts A7, B7--8, and C7--8 is noticeable particularly during task T2 (letter matching task). For this patient the brain area covered by these electrode contacts was associated with language processing (Wernicke's area). The observed higher level of synchronization probably reflects the higher degree of collectivity among neuronal assemblies that are involved in the cognitive operations comprising this task. This suggests that the eigenvalue decomposition method is capable of detecting short time changes of synchronization patterns associated with physiological processes in the human brain.

\section{Conclusion}
In this paper we have studied two multivariate phase synchronization analysis methods, namely a mean field approach \cite{Allefeld04} and an eigenvalue decomposition approach \cite{Allefeld05a}. While the mean field approach assumes the existence of a single synchronization cluster in the data, the eigenvalue decomposition approach appears to be capable of identifying multiple clusters. Based on the results of numerical simulations of multiple synchronization clusters within a lattice of 32 coupled identical Lorenz systems, we demonstrated that the mean field approach appears to be relatively robust even in situations where the single-cluster assumption is not entirely fulfilled. The eigenvalue decomposition approach successfully identified multiple synchronization clusters in our simulations and appears to be sensitive to detect clusters even for weak couplings. However, in case of non-vanishing inter-cluster synchronization the method failed to correctly assign the systems to their clusters in certain configurations. The influence of measurement noise, which was not discussed in depth here, can be expected to lead to the fragmentation of clusters present in the data. Nevertheless, the eigenvalue decomposition approach can be regarded as superior when analyzing data exhibiting different synchronization clusters.

When being applied to field data a number of influencing factors limit the significance of the eigenvalue decomposition approach and need further investigations. These factors include the finite length of available data, random spatio-temporal correlations, or even non-random correlations being induced by the data acquisition system (e.g. filtering or, as in the case of EEG recordings, the choice of a suitable reference).
Nevertheless, our preliminary applications of the eigenvalue decomposition approach to multichannel EEG recordings from epilepsy patients indicate that the method allows to gain deeper insights into the collective dynamics of neuronal networks, both under physiological and pathophysiological conditions. Despite the limited EEG database used in this study, the achieved results can be regarded as promising. Further evaluations on a larger EEG database are currently underway.

\begin{acknowledgments}
We are grateful to Carsten Allefeld, Anton Chernihovskyi, Andy M\"uller, Hannes Osterhage, Matth\"aus Staniek, and J\"org Wellmer for useful discussions and valuable comments. This work was supported by the Deutsche Forschungsgemeinschaft (Grant No. SFB-TR3 sub-project A2). S.B. was supported by a fellowship of the German National Academic Foundation (Studien\-stiftung).
\end{acknowledgments}

\end{document}